\pdfoutput=1
\documentclass[conference]{IEEEtran}
\IEEEoverridecommandlockouts

\usepackage{cite}
\usepackage{amsmath,amssymb,amsfonts}
\usepackage{graphicx}
\usepackage{booktabs}
\usepackage{textcomp}
\usepackage{xcolor}
\usepackage{url}
\usepackage{hyperref}
\hypersetup{hidelinks}

\usepackage{eso-pic}
\newcommand{\arxivnotice}{%
  \parbox{\textwidth}{\scriptsize\centering
  Accepted to the TrustAI-BioD Workshop at the 2026 IEEE International Conference on Bioinformatics and Biomedicine (BIBM).
  This is the author's accepted manuscript.\\[1pt]
  \copyright~2026 IEEE. Personal use of this material is permitted. Permission from IEEE must be obtained for all other uses,
  in any current or future media, including reprinting/republishing this material for advertising or promotional purposes,
  creating new collective works, for resale or redistribution to servers or lists, or reuse of any copyrighted component
  of this work in other works.}}
\AddToShipoutPictureBG*{%
  \AtPageLowerLeft{%
    \put(\LenToUnit{\dimexpr(\paperwidth-\textwidth)/2\relax},\LenToUnit{0.22in}){\arxivnotice}}}

\begin{document}

\title{Separating Genetic and Protein Evidence\\
in Open Targets: An Auditable Re-ranking Study}

\author{\IEEEauthorblockN{Ye Ye}
\IEEEauthorblockA{\textit{Independent Researcher}\\
New York, NY, USA\\
Email: yy4820@nyu.edu}}

\maketitle

\begin{abstract}
Biomedical target-ranking platforms combine genetics, literature, experimental evidence, tractability, and clinical precedent because each is useful for target discovery. A single integrated score, however, can be awkward to reuse for a narrower question such as which genes have direct evidence for disease risk or etiology. We examine this issue in Open Targets associations for Alzheimer disease (AD) and Parkinson disease (PD). Genetics-linked evidence and protein-oriented evidence are scored in separate channels and combined only at the end. In AD, the Open Targets overall top 20 contains 7 curated genetics-supported genes and 10 symptomatic pharmacology controls; the default late fusion contains 15 curated genes and no symptomatic controls. In PD, late fusion retains 13 curated genes and removes the 5 symptomatic controls present in the platform top 20. A genetics-only ranking reaches similar curated recall (14 genes in AD, 13 in PD), so most of the gain over the platform score comes from emphasizing genetic evidence; the protein channel mainly raises the number of tractable targets in the shortlist (from 9 to 12 in AD and from 8 to 10 in PD) without readmitting symptomatic controls. These results are stable across fusion weights from 0.35 to 0.75, perturbed protein-channel weights, and rule thresholds. An optional post-ranking rule further lowers high-clinical/low-genetic targets such as ACHE and GRIN1 without changing top-20 curated recall. The procedure is deterministic and uses public Open Targets fields. Because the component scores remain available, individual rank changes can be inspected and the ranking can be reproduced from the same data snapshot.
\end{abstract}


\section{Introduction}

Open Targets combines genome-wide associations, literature, expression, animal models, tractability, and clinical evidence for target--disease pairs~\cite{ochoa2023ot,koscielny2017ot}. The resulting overall score is useful for broad target exploration, but its interpretation depends on the question being asked. In particular, a high integrated score is not necessarily the same thing as strong genetic support for disease risk or etiology.

Alzheimer disease provides a simple example. APP, PSEN1, PSEN2, APOE, and TREM2 have substantial genetic support~\cite{karch2015alz,long2019alzheimer,guerreiro2013trem2}. Cholinesterases and NMDA receptors, by contrast, are prominent because they are established pharmacological mechanisms. They are important therapeutic targets, but high clinical or tractability evidence does not by itself make them strong candidates for disease etiology. If those evidence types are compressed into one scalar, a user can no longer tell whether a high rank came from genetics, clinical precedence, or druggability.

We treat this as an evidence-integration problem. Genetics-linked evidence and protein-oriented evidence are scored separately, their disagreement is retained, and the two channels are combined only in a final scoring step. We also test a small, explicit post-ranking rule for targets with strong clinical evidence but weak genetics. The Open Targets overall score remains the baseline rather than something the method is intended to replace. For trustworthy biomedical AI, the aim is a ranking whose evidence sources can be inspected and whose results can be regenerated from the same inputs.

The study uses a deterministic two-channel re-ranking on AD and then repeats the same procedure on Parkinson disease (PD), whose symptomatic pharmacology is different. We report the component scores, several fusion weights, rank changes, and the limits of the curated evaluation sets. This setup is intentionally simple: the main question is whether keeping evidence types separate makes a genetics-oriented shortlist easier to inspect without discarding non-genetic evidence altogether. We do not learn target labels or train a predictor, and we evaluate against both the platform ranking and a genetics-only ranking. The contribution is a reproducible evidence-accounting procedure, not a claim of superior causal target discovery.

\section{Related Work and Positioning}

Open Targets is a widely used integration of genetic, experimental, literature, tractability, and clinical evidence for therapeutic target identification~\cite{ochoa2023ot,koscielny2017ot}. Its overall association score is deliberately broad. We examine a narrower use case: reusing that broad score as a genetics-oriented ranking.

Genetics-first target identification has long been motivated by the higher clinical success observed for targets with human genetic support~\cite{nelson2015genetics,king2019genetic,minikel2024genetics}. For neurodegenerative disease, association studies and rare-variant findings have produced sets of well-supported risk genes, including TREM2 and many microglial or endolysosomal genes in AD~\cite{guerreiro2013trem2,bellenguez2022new}, and Mendelian or high-confidence risk genes in PD~\cite{blauwendraat2020genetics}. These studies motivate our curated evaluation sets, but they do not provide a complete gold standard.

Network-based target and drug prioritization provides a related view of the distinction between disease biology and established pharmacology. Emig et al. integrated disease and drug networks for target prediction~\cite{emig2013drug}, while Guney et al. used network proximity to distinguish palliative from disease-relevant drug effects and discussed this issue explicitly for Parkinson disease~\cite{guney2016network}. Our analysis uses only datatype scores already exposed by Open Targets. That makes it less expressive than a network model, but it also leaves the contribution of each evidence channel visible.

Genetics-only gene prioritization is a further point of comparison. Locus-to-gene models, for example, assign likely causal genes to GWAS loci~\cite{mountjoy2021l2g}, and Open Targets exposes the result as part of its genetic-association evidence. Our approach differs from existing prioritization methods in three ways. First, it does not learn weights from clinical outcomes or training labels; every weight and threshold is written down and can be changed by the user. Second, it keeps genetic and protein-oriented evidence as separate, reported channels instead of returning a single integrated or learned score. Third, it is tied to one explicitly stated task, nominating genes with genetic support for disease etiology, and is evaluated against controls defined for that task. A genetics-only ranking is therefore the most direct baseline, and we report it alongside the fused ranking.

\section{Problem Setup}
\label{sec:setup}

Let $\mathcal{G}_d$ be the targets associated with disease $d$ on Open Targets, truncated to the top 150 by the platform overall score. Each gene $g\in\mathcal{G}_d$ has datatype scores $s_{\tau}(g)\in[0,1]$ for genetic association, genetics-linked literature, general literature, RNA expression, animal models, clinical evidence, and related evidence types, together with binary tractability flags.

We consider a narrow task: re-rank this platform shortlist so that genes with established genetic support for the disease appear early, while genes that are prominent mainly because of symptomatic pharmacology appear later. The latter are not ``false'' drug targets. They serve as controls for a common task mismatch: therapeutic relevance being read as evidence of disease etiology.

Two evaluation sets are defined in advance from reviews and disease-genetics literature rather than from the ranking scores used below.

\paragraph{Curated genetics-supported genes.}
For AD we use 24 genes with convergent genetic support: APP, PSEN1, PSEN2, APOE, TREM2, SORL1, CLU, CR1, PICALM, BIN1, ABCA7, ADAM10, CD33, SPI1, PLCG2, MS4A4A, MS4A6A, EPHA1, CD2AP, INPP5D, ACE, GRN, TMEM106B, ABI3~\cite{karch2015alz,guerreiro2013trem2,bellenguez2022new}. Of these, 21 occur in the top-150 evaluation universe. For PD we use Mendelian and high-confidence risk genes (SNCA, LRRK2, GBA1, PRKN, PINK1, PARK7, VPS35, ATP13A2, DNAJC6, SYNJ1, FBXO7, PLA2G6, MAPT, VPS13C, CHCHD2); all 15 occur in the top 150~\cite{blauwendraat2020genetics}.

\paragraph{Symptomatic pharmacology controls.}
For AD we use ACHE, BCHE, GRIN1/2A/2B/2C/2D/3A/3B, SLC6A3, and SLC6A2 ($n{=}11$, all in the evaluation universe). For PD we use DRD1/2/3, SLC6A3, COMT, MAOA/B, DDC, ADORA2A, HTR2A, plus GRIN1/2B when present ($n{=}10$ in the evaluation universe). These proteins are legitimate therapeutic mechanisms. They are controls only for the genetics-oriented nomination task defined above.

The curated sets are incomplete and should not be read as exhaustive causal labels. Genes such as MPO, HFE, and ACE remain biologically contested despite appearing high in some rankings. Because the positive lists are defined by genetic evidence, a method that emphasizes genetic scores is aligned with the evaluation criterion by construction; gains over the broad platform score are therefore partly expected and are not independent confirmation of disease causality. We therefore report simple top-$k$ counts, mean ranks, and rank-based AUROC rather than treating this as a closed classification benchmark.

\section{Method}

\begin{figure}[t]
\centering
\includegraphics[width=0.95\columnwidth]{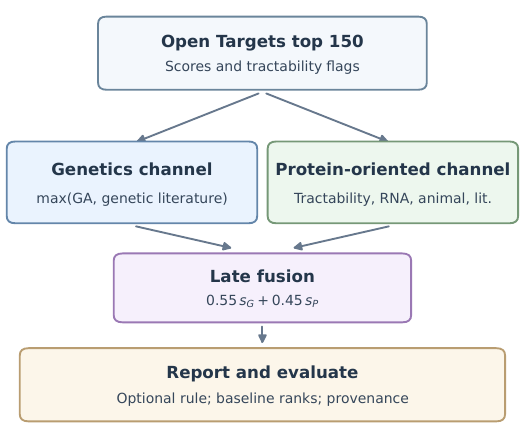}
\caption{Workflow. Evidence is split into two channels that are scored separately and fused only at the end; the optional rule is applied after fusion. Both channel scores are reported with the final ranking and compared with the platform and genetics-only baselines. No model is trained.}
\label{fig:workflow}
\end{figure}

Fig.~\ref{fig:workflow} summarizes the procedure.

\subsection{Separated evidence channels}
The genetics channel is
\begin{equation}
s_{\mathrm{G}}(g)=\max\big(s_{\mathrm{ga}}(g),\,s_{\mathrm{gl}}(g)\big),
\end{equation}
where $s_{\mathrm{ga}}$ and $s_{\mathrm{gl}}$ are the Open Targets genetic-association and genetics-linked literature scores. We use the maximum rather than the mean so that strong association evidence is not diluted by a sparse secondary channel.

The protein-oriented channel does not reuse the raw Open Targets pathway score. In the AD data, that field is dominated by a small set of CDK and mitochondrial genes and does not help distinguish the evaluation groups. We instead define
\begin{equation}
\begin{aligned}
s_{\mathrm{P}}(g)={}&0.35\,t(g)+0.25\,s_{\mathrm{rna}}(g)\\
&+0.20\,s_{\mathrm{an}}(g)+0.20\min\big(s_{\mathrm{lit}}(g),0.9\big).
\end{aligned}
\end{equation}
where $t(g)\in[0,1]$ is a saturated sum of tractability flags (approved drug, clinical stage, ligand-bound structure, high-quality pocket or ligand, druggable family, and antibody tractability). The literature term is capped at 0.9 so that citation volume alone cannot saturate the channel. The weights are fixed heuristics rather than fitted values. Tractability receives the largest weight because it is the component most specific to protein-level druggability; RNA expression, animal models, and literature are broader signals and receive smaller, similar weights. The channel is a protein-\emph{oriented} proxy, not a direct proteomics measurement, and we do not claim that these values are optimal. Section~\ref{sec:sens} shows that the results do not depend on these exact values.

\subsection{Late fusion}
The default combined score is
\begin{equation}
s_{\mathrm{F}}(g)=0.55\,s_{\mathrm{G}}(g)+0.45\,s_{\mathrm{P}}(g).
\end{equation}
The channels are scored before they are combined. This preserves provenance: for any highly ranked target, a reader can see whether its position is primarily genetic, protein-oriented, or supported by both. The default tilts the score slightly toward genetics, in line with the task, while leaving the protein channel a substantial share. It was not chosen as the best-performing setting: equal weights give a higher AD curated count. We also report equal weights $(0.5,0.5)$, genetics-heavy $(0.7,0.3)$, and protein-heavy $(0.3,0.7)$, and a full weight sweep in Section~\ref{sec:sens}.

\subsection{Explicit post-ranking rule}
We apply three explicit rules to the fused score:
\begin{enumerate}
\item high clinical evidence and weak genetics ($s_{\mathrm{clin}}\ge 0.8$ and $s_{\mathrm{G}}<0.25$) $\Rightarrow$ multiply by $0.35$;
\item strong genetics with little protein support ($s_{\mathrm{G}}\ge 0.45$ and $s_{\mathrm{P}}<0.12$) $\Rightarrow$ multiply by $0.85$;
\item protein/literature support without genetics ($s_{\mathrm{P}}\ge 0.45$, $s_{\mathrm{G}}<0.15$, $s_{\mathrm{clin}}<0.2$) $\Rightarrow$ multiply by $0.7$.
\end{enumerate}
The first rule addresses the task mismatch of interest. The other two keep the rule from collapsing into an implicit ``genetics only'' policy. The thresholds follow the semantics of the scores rather than a fit to the evaluation sets. A clinical score of at least 0.8 corresponds to strong clinical precedence for the indication; $s_{\mathrm{G}}<0.25$ lies below the 0.3 level used to count genetic support in the evaluation; and a multiplier of 0.35 is large enough to move such a target out of the top 20 without deleting it. Rules~2 and~3 apply milder penalties (0.85 and 0.70) to one-sided evidence, and their cutoffs are likewise screening heuristics rather than validated decision boundaries. In both diseases, rule~1 fires only on pharmacological targets (11 genes in AD, 34 in PD, mostly receptors and ion channels) and on no curated gene. All thresholds and multipliers are fixed and reported explicitly; they can be removed or changed without retraining a model, and Section~\ref{sec:sens} varies them.

\subsection{Why keep the channels separate}
For this task, auditability comes from retaining disagreement between evidence sources. On AD, the genetics and protein-channel top-20 lists share only six genes. Those disagreements are exactly the cases in which a single integrated score is hardest to interpret. Keeping the component scores available shows whether a high rank reflects disease genetics, tractability, experimental support, prior clinical use, or some combination of them.

\section{Study Design}

AD associations were retrieved from the Open Targets Platform GraphQL API for \texttt{MONDO\_0004975} on 18 September 2026 (top 150 rows). PD associations used \texttt{MONDO\_0005180} with the same query shape. No patient-level data are used.

We compare literature-only ranking ($s_{\mathrm{lit}}$), the Open Targets overall score, genetic association alone, the genetics channel, the protein channel, four fusion weights, and fusion plus the post-ranking rule. At $k{=}20$, we report the number of curated genetics-supported genes, the number of symptomatic controls, and the number of genes with $s_{\mathrm{G}}\ge 0.3$. We also report rank-based AUROC for each evaluation set and the mean rank of its members.

The evaluation universe is conditioned on the Open Targets top 150. It is therefore a re-ranking experiment, not a whole-genome target screen. The truncation can exclude genes that would rank highly under a genetics-only query, so conclusions are limited to ordering within this preselected platform shortlist.

All tables and figures are generated from cached GraphQL responses by a deterministic scoring script. Once the cached input is fixed, repeated runs produce the same ranking; there is no stochastic training step. The cached JSON responses, derived CSV files, and scoring scripts are publicly available (see Data and code).

We use the same stored inputs for every comparison and report multiple fusion weights rather than selecting a setting from the best AD result. PD is used as a second disease with a different symptomatic-control set, not as a held-out estimate of generalization.

\section{Results}

\subsection{Alzheimer disease}
Table~\ref{tab:ad} summarizes the AD rankings. The Open Targets overall score places GRIN1 5th and ACHE 10th; 10 of its top 20 entries are symptomatic controls. Literature-only ranking contains 13 curated genes and 1 control. The genetics channel recovers 14 curated genes with no symptomatic controls, including PSEN1, APP, CR1, PSEN2, APOE, TREM2, PLCG2, MS4A6A, EPHA1, BIN1, CD33, and ADAM10. The protein channel, even after removing the pathway field, still places GRIN2B, GRIN1, and ACHE near the top.

\begin{table}[t]
\centering
\caption{AD shortlists inside the Open Targets top 150. Higher curated counts and lower symptomatic-control counts are preferred for the task defined in Section~\ref{sec:setup}. Genetic support counts genes with $s_{\mathrm{G}}\ge 0.3$.}
\label{tab:ad}
\setlength{\tabcolsep}{3.2pt}
\begin{tabular}{lccc}
\toprule
Method & Curated@20 & Sympt.@20 & Genetic@20 \\
\midrule
Literature only & 13 & 1 & 14 \\
Open Targets overall & 7 & 10 & 9 \\
Genetic association only & 14 & 0 & 20 \\
Genetics channel & 14 & 0 & 20 \\
Protein channel & 5 & 7 & 6 \\
Equal fusion $0.5/0.5$ & \textbf{16} & \textbf{0} & \textbf{20} \\
Genetics-heavy $0.7/0.3$ & 15 & 0 & 20 \\
Protein-heavy $0.3/0.7$ & 14 & 2 & 18 \\
Default fusion $0.55/0.45$ & 15 & 0 & 20 \\
Fusion + rule & 15 & 0 & 20 \\
\bottomrule
\end{tabular}
\end{table}

Equal weights give the highest curated count at $k{=}20$ on this label set (16 genes).
Protein-heavy fusion is the only fused variant that lets symptomatic controls back in (2 in the top 20).
Default fusion and the post-ranking rule have the same top-20 counts. We therefore report the full weight sensitivity rather than selecting a single setting from this disease alone.

\begin{figure}[t]
\centering
\includegraphics[width=\columnwidth]{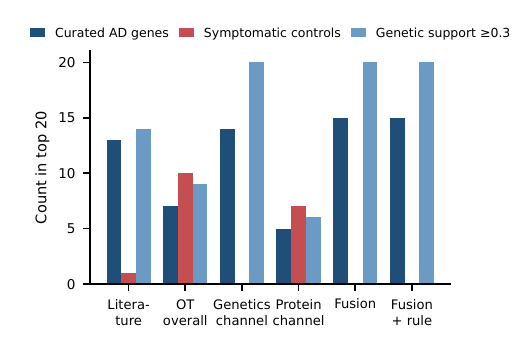}
\caption{Composition of the AD top 20 under each ranking. Late fusion removes symptomatic controls from the top 20 while retaining strong genetic support.}
\label{fig:metrics}
\end{figure}

\begin{figure}[t]
\centering
\includegraphics[width=0.92\columnwidth]{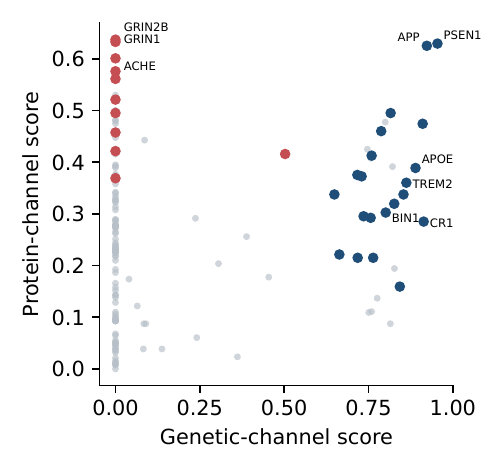}
\caption{Genetic versus protein scores on 150 AD associations. Dark points are curated genetics-supported genes; red points are symptomatic pharmacology controls.}
\label{fig:scatter}
\end{figure}

Fig.~\ref{fig:metrics} summarizes the same counts. Fig.~\ref{fig:scatter} shows the channel separation directly: most curated genes have high genetics scores, whereas the symptomatic controls cluster at lower genetics scores and higher protein-oriented scores. The two channel-specific top-20 lists overlap in only ACE, APP, CD33, MPO, PSEN1, and PSEN2. TREM2, CR1, PLCG2, and BIN1 are prominent only in the genetics list, while ACHE, several GRIN receptors, CDK5, MTOR, and GSK3B are prominent only in the protein list.

\subsection{Effect of the post-ranking rule}
\begin{figure}[t]
\centering
\includegraphics[width=0.92\columnwidth]{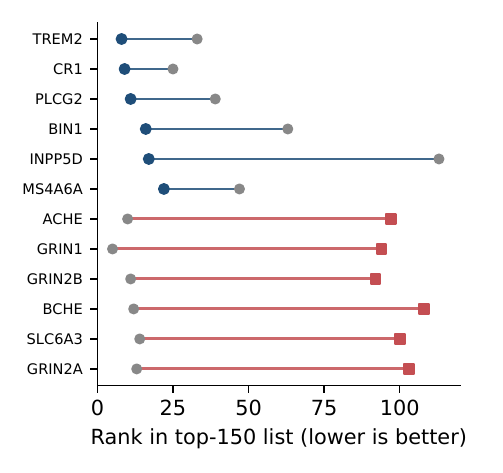}
\caption{Rank movement from the Open Targets overall score (gray) to fused scoring with the post-ranking rule (color) for selected AD genes. Lower rank is better. Genetics-supported genes move toward the top; symptomatic cholinesterase and NMDA-receptor controls move downward.}
\label{fig:ranks}
\end{figure}

Late fusion already removes every AD symptomatic control from the top 20. The post-ranking rule mainly changes their positions farther down the list (Fig.~\ref{fig:ranks}). Table~\ref{tab:shift} gives representative rank shifts.
ACHE falls from overall rank 10 to rule-adjusted rank 97; GRIN1 from 5 to 94; GRIN2B from 11 to 92.
TREM2 moves from 33 to 8, CR1 from 25 to 9, PLCG2 from 39 to 11, BIN1 from 63 to 16, and INPP5D from 113 to 17.

The component scores explain these moves. TREM2 ($s_{\mathrm{G}}{=}0.86$, $s_{\mathrm{P}}{=}0.36$) and INPP5D ($0.73$, $0.37$) have strong genetic evidence but little clinical or tractability evidence, which keeps them low in the platform score. ACHE and GRIN1 have no genetic evidence ($s_{\mathrm{G}}{=}0$) but maximal clinical scores and high protein scores (0.60 and 0.63); fusion alone moves them to ranks 36 and 35, and the rule moves them below 90. The same scores also show what the protein channel adds and costs. ACE ($s_{\mathrm{P}}{=}0.49$) rises from 12th in the genetics channel to 4th, and MPO (0.48), which is not in the curated set and remains debated, rises from 15th to 6th; HFE, with weak protein support (0.19), falls from 9th to 20th.

\begin{table}[t]
\centering
\caption{Selected AD rank changes inside the top-150 universe. Lower is better.}
\label{tab:shift}
\setlength{\tabcolsep}{4pt}
\begin{tabular}{lrrr}
\toprule
Gene & Overall & Fusion & Rule \\
\midrule
TREM2 & 33 & 8 & 8 \\
CR1 & 25 & 9 & 9 \\
PLCG2 & 39 & 11 & 11 \\
BIN1 & 63 & 16 & 16 \\
INPP5D & 113 & 17 & 17 \\
ACHE & 10 & 36 & 97 \\
GRIN1 & 5 & 35 & 94 \\
GRIN2B & 11 & 34 & 92 \\
BCHE & 12 & 46 & 108 \\
SLC6A3 & 14 & 39 & 100 \\
\bottomrule
\end{tabular}
\end{table}

Mean ranks tell the same story without a $k$ cutoff.
Under the overall score, curated AD genes have mean rank 39.4 and symptomatic controls 14.1. After the post-ranking rule, the corresponding means are 13.8 and 97.3. The rank AUROC for curated genes rises from 0.780 to 0.979; the AUROC for symptomatic controls, treated as the positive class for this diagnostic calculation, falls from 0.942 to 0.343. A genetics-only ranking already gives mean ranks of 14.9 and 84.7, so relative to that baseline the rule mainly pushes controls lower rather than improving the curated genes. All of these AUROC values are computed against genetics-defined labels and should be read with that dependency in mind.

\subsection{Parkinson disease as a replication}
\label{sec:pd}
PD provides a different symptomatic-control structure. Its prominent therapeutic mechanisms are dopaminergic and catecholamine related (for example DRD2, COMT, MAOB, and DDC) rather than cholinesterase inhibition, although GRIN subunits also appear.

Table~\ref{tab:pd} shows a milder version of the same task mismatch. The overall score already contains 13 curated genes in its top 20, but it also contains 5 symptomatic controls. The protein channel contains 3 curated genes and 7 controls, including GRIN2B and DRD2. The genetics channel and late fusion retain 13 curated genes with no symptomatic controls. As in AD, the post-ranking rule primarily changes the mean rank of the control set (23.2 under the overall score and 100.8 after the post-ranking rule) rather than the top-20 curated count.

\begin{table}[t]
\centering
\caption{PD shortlists inside the Open Targets top 150.}
\label{tab:pd}
\setlength{\tabcolsep}{3.2pt}
\begin{tabular}{lccc}
\toprule
Method & Curated@20 & Sympt.@20 & Genetic@20 \\
\midrule
Literature only & 10 & 4 & 14 \\
Open Targets overall & 13 & 5 & 14 \\
Genetics channel & \textbf{13} & \textbf{0} & \textbf{20} \\
Protein channel & 3 & 7 & 5 \\
Default fusion & \textbf{13} & \textbf{0} & \textbf{20} \\
Fusion + rule & \textbf{13} & \textbf{0} & \textbf{20} \\
\bottomrule
\end{tabular}
\end{table}

The PD protein-only top 20 includes ACHE, MAPT, GRIN2B, GRIN2A, DRD2, ADORA2A, COMT, and MAOB.
The genetic-only top 20 is almost entirely the Mendelian and GWAS set: PLA2G6, PRKN, PINK1, ATP13A2, DNAJC6, LRRK2, SNCA, PARK7, FBXO7, GBA1, VPS35.
The same channel disagreement is therefore visible in a disease with a different therapeutic profile. In PD the protein channel promotes MAPT (genetics rank 15, fused rank 3) and the non-curated UCHL1 (39 to 16), PSAP (30 to 12), and NR4A2; none of these promotions makes a gene a validated causal target. Conversely, DRD2 is 5th in the protein channel but falls from overall rank 17 to fused rank 44 and rule-adjusted rank 89. The protein channel also has a cost: VPS13C, a curated gene with almost no protein-oriented evidence ($s_{\mathrm{P}}{=}0.08$), falls from 13th in the genetics channel to 28th.

\subsection{Genetics-only baseline and protein-channel ablation}
\label{sec:ablation}
Because the curated sets consist of genetics-supported genes, part of any improvement over the platform score is expected to come simply from emphasizing genetic evidence. The fair comparison for the fused ranking is therefore a genetics-only ranking. Table~\ref{tab:ablation} reports it together with an ablation that removes one component of the protein channel at a time. T@20 counts top-20 genes with tractability score $t(g)\ge 0.5$.

\begin{table}[t]
\centering
\caption{Genetics-only baseline and protein-channel ablation (default 0.55/0.45 fusion). C@20: curated genes in the top 20; AUROC: rank AUROC for curated genes; T@20: top-20 genes with tractability $t(g)\ge 0.5$. No row places a symptomatic control in the top 20.}
\label{tab:ablation}
\setlength{\tabcolsep}{3pt}
\begin{tabular}{lcccccc}
\toprule
 & \multicolumn{3}{c}{AD} & \multicolumn{3}{c}{PD} \\
\cmidrule(lr){2-4}\cmidrule(lr){5-7}
Ranking & C@20 & AUROC & T@20 & C@20 & AUROC & T@20 \\
\midrule
Genetic assoc.\ only & 14 & 0.970 & 9 & 12 & 0.976 & 8 \\
Genetics channel & 14 & 0.970 & 9 & 13 & 0.984 & 8 \\
Default fusion & 15 & 0.978 & 12 & 13 & 0.979 & 10 \\
\ \ $-$ tractability & 17 & 0.985 & 9 & 13 & 0.989 & 9 \\
\ \ $-$ RNA expression & 15 & 0.976 & 12 & 12 & 0.978 & 10 \\
\ \ $-$ animal model & 16 & 0.978 & 12 & 12 & 0.976 & 9 \\
\ \ $-$ literature & 14 & 0.968 & 12 & 12 & 0.975 & 11 \\
\bottomrule
\end{tabular}
\end{table}

Fusion and the genetics channel reach similar curated recall. In AD, fusion adds one curated gene at $k{=}20$ and raises the curated AUROC from 0.970 to 0.978; in PD, the curated count is unchanged and the AUROC is slightly lower (0.979 versus 0.984). The small net changes hide larger changes in membership: in each disease, fusion replaces five of the 20 genetics-channel genes. In AD, the curated GRN, INPP5D, and SORL1 enter together with ABCA1 and CTSB, while the curated CD2AP and MS4A6A leave; in PD, the curated SYNJ1 enters and the curated VPS13C leaves. The protein channel therefore does not materially improve recovery of genetics-supported genes. What it changes is the composition of the shortlist: the number of tractable top-20 genes rises from 9 to 12 in AD and from 8 to 10 in PD, while no symptomatic control enters. The ablation shows where this trade-off comes from. Removing tractability gives the highest curated recall (17 in AD) but returns the tractable count to the genetics-only level; removing literature lowers curated recall in both diseases. The protein channel is thus best read as a way to reorder an already genetics-supported shortlist toward targets that are easier to act on, not as a source of additional genetic signal.

\subsection{Sensitivity to weights and thresholds}
\label{sec:sens}
\begin{table}[t]
\centering
\caption{Fusion-weight sensitivity. Each cell gives curated genes / symptomatic controls in the top 20; $w$ is the genetics weight in $w\,s_{\mathrm{G}}+(1-w)\,s_{\mathrm{P}}$.}
\label{tab:weights}
\setlength{\tabcolsep}{4pt}
\begin{tabular}{lcc}
\toprule
Genetics weight $w$ & AD & PD \\
\midrule
0.30 & 14 / 2 & 12 / 0 \\
0.40 & 16 / 0 & 12 / 0 \\
0.50 & 16 / 0 & 13 / 0 \\
0.55 (default) & 15 / 0 & 13 / 0 \\
0.70 & 15 / 0 & 12 / 0 \\
0.85 & 14 / 0 & 13 / 0 \\
1.00 (genetics only) & 14 / 0 & 13 / 0 \\
\bottomrule
\end{tabular}
\end{table}

\emph{Fusion weight.} Table~\ref{tab:weights} shows selected weights; the full sweep from 0 to 1 in steps of 0.05 gives the same picture. The response is not monotonic, and this is a sensitivity check, not a hyperparameter search. Across the sweep, no symptomatic control enters the top 20 once the weight reaches 0.35 in AD or 0.30 in PD. Between 0.35 and 0.75, the AD curated count stays between 15 and 16 and the PD count between 12 and 13. The default 0.55 lies inside this plateau in both diseases; the results do not hinge on it.

\emph{Protein-channel weights.} Over 1{,}000 random weight vectors drawn around the default (Dirichlet, concentration 40) and over a grid that scales each weight by 0.5, 1, or 1.5, the AD curated count ranges from 14 to 16 and the PD count from 11 to 14. A single symptomatic control enters the AD top 20 in 0.3\% of the random draws and in 2 of the 81 grid settings, and never in PD. Changing the literature cap to 0.8 or 1.0 changes the curated count by at most one gene.

\emph{Post-ranking rule.} Across 36 settings (clinical threshold 0.6--0.9, genetics threshold 0.15--0.35, multiplier 0.2--0.5), top-20 counts and curated AUROC are unchanged in both diseases. Only the mean rank of the symptomatic controls varies (74--110 in AD and 77--116 in PD), consistent with the rule acting below the top 20.

\subsection{Scope of the experiment}
The experiment is deliberately narrow. It cannot discover genes outside the Open Targets top 150, and it does not replace structure-based, single-cell, or functional validation. A high fused rank is not a claim that a gene is a true therapeutic target: MPO and HFE, for example, rise under AD fusion while remaining debated. The two-channel fusion uses only datatype fields that Open Targets provides for every disease, so it can be run elsewhere without modification, but we have tested it only on two neurodegenerative diseases. Other disease categories would change both the channels and the evaluation. In oncology, somatic mutation may be the relevant genetic signal and would belong in the genetics channel. In monogenic disorders, the genetics channel alone may already be decisive. In infectious disease, pathogen-specific evidence may matter more than human germline genetics. In diseases where approved drugs act on causal drivers, such as many metabolic and inflammatory disorders, the symptomatic-control concept and the post-ranking rule would need to be redefined or dropped. In each case the curated and control sets would have to be rebuilt for the disease, and an independently labeled test set or prospective validation would be needed.

\section{Discussion}

These experiments expose a mismatch between what a score summarizes and what a user asks it to mean. Open Targets combines several evidence types for good reasons, and its overall score is useful for general target exploration. Clinical precedence, tractability, and human genetics, however, are not interchangeable evidence for disease etiology. Reusing a broad score for a genetics-oriented discovery question can therefore produce a ranking that is difficult to interpret even when the underlying data are sound.

The separated channels make the source of a high rank visible. A candidate can be supported mainly by genetics, mainly by protein-oriented evidence, or by both. The fusion is simple enough to recompute directly, and the post-ranking rule is written out rather than learned implicitly. This makes it possible to inspect a rank change, reproduce it from the same data snapshot, and see where the method should not be applied.

The protein-oriented channel adds little to the recovery of genetics-supported genes; a genetics-only ranking performs about as well on that measure (Section~\ref{sec:ablation}). Its contribution is different: it brings tractability, expression, and experimental-model evidence into the ordering, so that the shortlist contains more targets that can be acted on. Keeping that information in a separate channel allows it to affect the final score without being interpreted as genetic evidence, and the cost is visible, as with VPS13C in PD. No downstream validation shows that these membership changes improve therapeutic success, and because the evaluation labels favor genetic evidence, they are not an independent test of the protein channel.

The post-ranking rule is the most disease-specific part of the method. In AD and PD, high clinical evidence with weak genetics often marks symptomatic pharmacology. That relationship need not hold elsewhere. In a disease area where successful therapies directly target causal drivers, the same rule could lower exactly the targets one would want to keep. We therefore report it as an optional adjustment, separate from the two-channel fusion, and do not treat it as a generally transferable rule.

\section{Limitations}

This is a retrospective study of two diseases and one public knowledge graph. The curated genetics sets are literature-derived and incomplete, and the symptomatic-control sets reflect a deliberately narrow evaluation question rather than a universal definition of therapeutic relevance. The top-150 evaluation universe is selected by the Open Targets overall score, which limits conclusions about whole-genome ranking behavior. The channel weights and post-ranking thresholds are fixed heuristics rather than parameters learned or validated on a held-out disease. We report weight and threshold sensitivity, but two diseases are not enough to establish generalization or to support a formal cross-disease statistical claim.

The protein channel is also only an approximation to protein-level evidence. It uses tractability flags, RNA expression, animal models, and literature rather than raw proteomics, structural measurements, or single-cell atlases. Finally, the study evaluates ranking behavior, not prospective target validation. A better position in these lists should be read as better alignment with the stated evidence semantics, not proof that the gene will produce a successful therapy.

\section{Conclusion}

Integrated biomedical scores can be useful for broad exploration while still being a poor fit for a narrower ranking question. In these AD and PD examples, keeping genetics-linked and protein-oriented evidence separate removes symptomatic pharmacology controls from the top 20 of a genetics-oriented shortlist while retaining the non-genetic channel for later fusion. Most of this gain is also achieved by a genetics-only ranking; the protein channel mainly shifts the shortlist toward tractable targets, and the results are stable across the weights and thresholds we tested. The optional rule changes later ranks but does not improve top-20 curated recall. The useful property here is transparency: a reader can see which evidence moved a target and can regenerate the ranking from the same public-data snapshot.

\section*{Acknowledgment}
Generative AI tools (xAI Grok, OpenAI ChatGPT, and Anthropic Claude) were used for portions of code implementation, literature discovery, language and editorial revision, the camera-ready sensitivity analyses, and figure preparation. The author developed the research idea, study design, evaluation protocol, analysis, and interpretation, and reviewed and verified all AI-assisted outputs.

\section*{Data and code}
All AD and PD scores were derived from one Open Targets GraphQL query per disease. The cached JSON responses, derived CSV files, and the scoring and sensitivity-analysis scripts are available at \url{https://github.com/biubiutomato/opentargets-two-channel}. No patient-level data or stochastic model training are used.

\appendices
\footnotesize

\section{Alzheimer disease: top-20 lists}
\textbf{Open Targets overall:}
APP, PSEN1, PSEN2, APOE, GRIN1, SORL1, GRIN3B, CDK5, ADAM10, ACHE, GRIN2B, BCHE, GRIN2A, SLC6A3, GRIN2C, GRIN2D, CDK5R1, GRIN3A, ACE, HFE.

\textbf{Genetics channel:}
PSEN1, APP, CR1, PSEN2, APOE, TREM2, PLCG2, MS4A6A, HFE, EPHA1, APH1B, ACE, H2BC4, BIN1, MPO, CD33, NCK2, CD2AP, ADAM10, SHARPIN.

\textbf{Protein channel:}
GRIN2B, GRIN1, PSEN1, APP, ACHE, SLC6A3, GRIN2A, CDK5R1, GRIN2C, CDK5, GSK3B, BCHE, ACE, MTOR, SMPD1, PSAP, MPO, HMGCR, PSEN2, CD33.

\textbf{Default fusion / post-ranking rule (identical at $k{=}20$):}
PSEN1, APP, PSEN2, ACE, APOE, MPO, CD33, TREM2, CR1, APH1B, PLCG2, ADAM10, ABCA1, CTSB, EPHA1, BIN1, INPP5D, GRN, SORL1, HFE.

\section{Parkinson disease: top-20 lists}
\textbf{Open Targets overall:}
LRRK2, ATP13A2, SNCA, PINK1, PRKN, DNAJC6, PARK7, FBXO7, VPS35, VPS13C, PLA2G6, GBA1, MAPT, GRIN1, CHRM1, PSAP, DRD2, DRD3, GRIN2B, COMT.

\textbf{Genetics channel:}
PLA2G6, PRKN, PINK1, ATP13A2, DNAJC6, LRRK2, SNCA, PARK7, FBXO7, GBA1, VPS35, BST1, VPS13C, ENSG00000288636, MAPT, PACRG, NR4A2, SH3GL2, TMEM175, RIT2.

\textbf{Protein channel:}
ACHE, MAPT, GRIN2B, GRIN2A, DRD2, UCHL1, SCN8A, GRIN1, GSK3B, COMT, MAOB, ADORA2A, MTOR, DDC, PSAP, SNCA, SMPD1, SCN1A, BCHE, GBA1.

\textbf{Default fusion / post-ranking rule (identical at $k{=}20$):}
PINK1, PLA2G6, MAPT, SNCA, LRRK2, PARK7, GBA1, PRKN, ATP13A2, NR4A2, BST1, PSAP, TMEM175, VPS35, DNAJC6, UCHL1, GPNMB, FBXO7, ITPKB, SYNJ1.

\end{document}